\documentclass[fleqn,twoside]{article}
\usepackage{espcrc2}
\def\lesssim{\mathrel{\mathpalette\vereq<}}

\makeatletter
\def\vereq#1#2{\lower3pt\vbox{\baselineskip1.5pt \lineskip1.5pt
\ialign{$\m@th#1\hfill##\hfil$\crcr#2\crcr\sim\crcr}}}
\makeatother

\usepackage{graphicx}

\newcommand{\AmS}{{\protect\the\textfont2
  A\kern-.1667em\lower.5ex\hbox{M}\kern-.125emS}}

\title{GUT, Neutrinos, and Baryogenesis}

\author{Hitoshi Murayama\address{Department of Physics,
    University of California\\
    Berkeley, CA 94720}%
  \address{Theoretical Physics Group,
    Lawrence Berkeley National Laboratory\\
    Berkeley, CA 94720}%
  \thanks{This work was supported in part by the DOE Contract DE-AC03-76SF00098
    and in part by the NSF grant PHY-0098840.}}
       
\begin{document}

\begin{abstract}
  It is an exciting time for flavor physics.  In this talk, I discuss
  recent topics in baryogenesis and leptogenesis in light of new data,
  and implications in $B$ and neutrino physics.  I also discuss
  current situation of grand unified theories concerning coupling
  unification, proton decay, and indirect consequences in lepton
  flavor violation and $B$ physics.  I explain attempts to understand
  the origin of flavor based on flavor symmetry, in particular
  ``anarchy'' in neutrinos.  \vspace{1pc}
\end{abstract}

\maketitle

\section{Introduction}

Flavor physics is going through a big revolution.  Neutrino
oscillation has been a big discovery, and new data on quark sector are
pouring in.  Given the excitement in flavor physics, I will discuss
recent topics on baryogenesis and grand unification, and their
connection to flavor physics in this talk.

\section{Baryogenesis}

Why there is only matter in Universe but no anti-matter is one of the
big questions in cosmology and particle physics.  Because of its
fundamental importance, it is often quoted as one of the primary
reasons to study CP violation in flavor physics.  

First of all, the amount of baryon we have in our Universe had been
quite well determined by Big-Bang Nucleosynthesis.  The only free
parameter in this theory is the amount of baryons that determines the
rate of nuclear fusion process in early Universe.  People often quote
the baryon-to-photon ratio $\eta = n_B/n_\gamma$, namely the ratio of
the baryon number over the number of photons in a fixed volume,
because this quantity does not change as the Universe
expands.\footnote{When a particle species freezes out, however, the
  ratio does change.  The baryon-to-entropy ratio is constant across
  the thresholds.}  There was a period when different determinations
of baryon-to-photon ratio did not agree with each other, and it was
said to be a crisis.  The problem has largely disappeared recently,
and there emerged a consensus on the baryon-to-photon ratio.  Let me
quote two quantitative analyses: 
\begin{equation}
  \label{eq:1}
  \eta = \frac{n_B}{n_\gamma} = \left\{ \begin{array}{ll}
      \left( 4.7^{+1.0}_{-0.8} \right) \times 10^{-10} 
      & \cite{BBN1} \\
      (5.0 \pm 0.5) \times 10^{-10}
      & \cite{BBN2}
    \end{array} \right. .
\end{equation}
This determination of the baryon-to-photon ratio is also consistent
with the analysis that combines power spectrum in cosmic microwave
anisotropy, cluster mass, and large scale structure \cite{CMB}.
Therefore, we know the baryon-to-photon ratio of the Universe with a
good confidence. 

When I mention the word ``anti-matter'' in a public talk, it causes a
certain level of fear among the audience.  It sounds something scary
to them.  And they are right.  When the Universe was hot, at a
temperature of 1~GeV, there were practically equal amount of matter
and anti-matter.  As the temperature decreased, anti-matter has
annihilated with matter, leaving only radiation.  However, at the
level of one out of ten billions or so, there was an excess in the
amount of matter over anti-matter, and this small excess is {\it
  us}\/.  We have survived {\it The Great Annihilation}\/.  This
realization immediately leads to a question: ``What caused a small
excess in the amount of matter over anti-matter?''  This excess is
called the baryon asymmetry of Universe.

Sakharov pointed out that the small baryon asymmetry may be understood
as a consequence of microphysics from a Universe with no asymmetry
only if three conditions are satisfied:
\begin{enumerate}
\item Existence of process that violates the baryon number.
\item CP Violation.
\item Departure from thermal equilibrium.
\end{enumerate}
The first requirement is obvious.  If the Universe had no asymmetry as
its initial condition, generation of baryon asymmetry is possible only
if the baryon number can change.  Then there may be a finite rate of a
process that increases the baryon number $\Gamma(\Delta B > 0) \neq
0$.  If, however, CP were an exact symmetry, a process and its CP
conjugate process would have the same rate.  Because the baryon number
is CP-odd, it would imply that $\Gamma (\Delta B > 0) = \Gamma (\Delta
B < 0)$, and no baryon asymmetry can be generated.  To make these
rates different, CP violation is mandatory.  Even so, thermal
equilibrium, by definition, has the same rates for a process and its
inverse process.  Similarly to the CP conservation, it would also
imply $\Gamma (\Delta B > 0) = \Gamma (\Delta B < 0)$, and no baryon
asymmetry would be generated.  Therefore, departure from thermal
equilibrium is necessary to generate the baryon asymmetry.

It was once hoped that grand unified theories (GUTs) would provide the
mechanism of generating baryon asymmetry, a.k.a. baryogenesis
\cite{Yoshimura}.  GUTs indeed necessarily break baryon number.  If a
heavy particle from GUTs remain in the early Universe after the
temperature drops below its production threshold, the leftover amount
exhibits the departure from thermal equilibrium.  Then their decay, if
CP is violated, may preferentially produce baryons over anti-baryons,
thereby generating baryon asymmetry.  It is encouraging that such a
decay asymmetry had been established, namely $\varepsilon' \neq 0$ or
equivalently $\Gamma(K^0 \rightarrow \pi^+ \pi^-) \neq
\Gamma(\overline{K}^0 \rightarrow \pi^+ \pi^-)$.

\begin{figure}[t]
  \begin{center}
    \includegraphics[width=0.6\columnwidth]{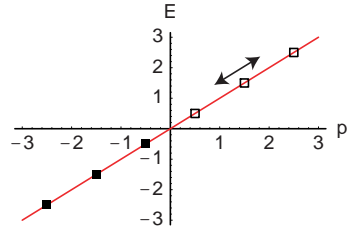}
    \caption[Dirac]{The energy levels of the Dirac equation in the
      presence of fluctuating $W$-field move up and down.  All
      negative energy states are occupied while the positive energy
      states vacant in the ``vacuum'' configuration.}
    \label{fig:Dirac}
  \end{center}
\end{figure}

\begin{figure}[t]
  \begin{center}
    \includegraphics[width=0.6\columnwidth]{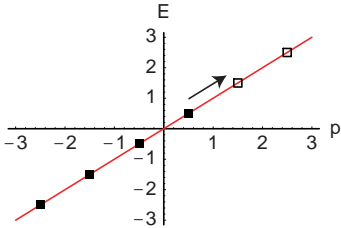}
    \caption[Dirac2]{Once in a while, the fluctuation in the $W$-field
      becomes so large that the energy levels of the Dirac equation in
      the presence of fluctuating $W$-field shift all the way by one
      unit.  Then a positive energy state is occupied and a particle
      is created.}
    \label{fig:Dirac2}
  \end{center}
\end{figure}

However, the effect of electroweak anomaly changed the picture quite
dramatically.  The Standard Model actually violates $B$
\cite{anomaly}.  In the Early Universe when the temperature was above
250~GeV, there was no Higgs boson condensate and $W$ and $Z$ bosons
were massless (so where all quarks and leptons).  Therefore $W$ and
$Z$ fields were just like electromagnetic field in the hot plasma and
were fluctuating thermally.  The quarks and leptons move around under
the fluctuating $W$-field background.  To see what they do, we solve
the Dirac equation for fermions coupled to $W$.  There are positive
energy states that are left vacant, and negative energy states that
are filled in the ``vacuum'' (Fig.~\ref{fig:Dirac}).  As the $W$-field
fluctuates, the energy levels fluctuate up and down accordingly.  Once
in a while, however, the fluctuation becomes so large that all energy
levels are shifted by one unit (Fig.~\ref{fig:Dirac2}).  Then you see
that one of the positive energy states is now occupied.  There is now
a particle! This process occurs in the exactly the same manner for
every particle species that couple to $W$, namely for all left-handed
lepton and quark doublets.  This effect is called the electroweak
anomaly.  Therefore the electroweak anomaly changes (per generation)
$\Delta L =1$, and $\Delta q = 1$ for all three colors, and hence
$\Delta B = 1$.  Note that $\Delta (B-L) = 0$; the electroweak anomaly
preserves $B-L$.

Because of this process, the pre-existing $B$ and $L$ are converted to
each other to find the chemical equilibrium at $B \sim 0.35 (B-L)$, $L
\sim -0.65 (B-L)$ \cite{equilibrium}. In particular, even if there was
both $B$ and $L$, both of them get washed out if $B-L$ was zero.

Given this problem, there are now two major directions in the
baryogenesis.  One is the electroweak baryogenesis \cite{KRS}, where
you try to generate $B=L$ at the time of the electroweak phase
transition so that they do not get washed out further by the
electroweak anomaly.  The other is the leptogenesis \cite{FY}, where
you try to generate $L\neq 0$ but no $B$ from neutrino physics well
before the electroweak phase transition, and $L$ gets partially
converted o $B$ due to the electroweak anomaly.

\subsection{Electroweak Baryogenesis}

It appears at the first sight that the baryogenesis is possible in the
Minimal Standard Model.  First, the baryon number is violated as we
discussed above.  Second, CP is also violated in the Standard Model.
Third, if the phase transition of electroweak symmetry breaking is
first order, the coexistence of broken and unbroken phases at the
phase transition is a departure from equilibrium.  Then all three
conditions by Sakharov are satisfied.  The question is if enough
baryon asymmetry can be generated quantitatively \cite{CKN}.  There are
at least two big problems in the Standard Model.  The first problem is
the order of phase transition.  The first order phase transition is
possible only if the Higgs boson is relatively light, $m_H \lesssim
60$~GeV.  Above this mass, the phase transition becomes second order,
and there is no departure from equilibrium.  The LEP bound on the
Higgs boson has excluded this possibility.  The other problem is the
size of CP violation.  In the Standard Model, any CP violating effects
must be proportional to the so-called Jarlskog parameter, $J =
\Im(\mbox{det}[M_u^\dagger M_u, M_d^\dagger M_d])$.  At the phase
transition temperature $T_{EW} \simeq v$, the dimensionless quantity
that characterizes the size of CP violation is $J/v^{12} \sim 10^{-20}
\ll 10^{-10}$.  Unless there is a mechanism of tremendous enhancement
by ten orders of magnitude, the resulting baryon asymmetry would be too
small.

\begin{figure}[t]
  \begin{center}
    \includegraphics[width=0.8\columnwidth]{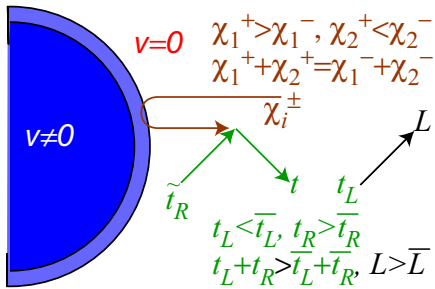}
    \caption[bubbles]{The mechanism of baryogenesis in the MSSM.  The
      bubbles of the true vacuum with broken electroweak symmetry $v
      \neq 0$ forms and expands into the false vacuum with unbroken
      electroweak symmetry.  The charginos bounce off the expanding
      bubble walls.  Because of the CP violation in the chargino mass
      matrix, the reflection probabilities are different for different
      charginos.  The interaction of light scalar top quark with the
      charginos convert the difference in the charginos to the
      asymmetry between left-handed and right-handed top quarks.  At
      this point, there is no overall top quark asymmetry and no
      baryon number.  Then the asymmetry in the left-handed top quark
      is partially converted to the lepton asymmetry due to the
      electroweak anomaly.  The reduced asymmetry in the left-handed
      top quark and the remaining (unaffected) asymmetry in the
      right-handed top quark no longer cancel and there is net baryon
      asymmetry. }
    \label{fig:bubbles}
  \end{center}
\end{figure}

The Minimal Supersymmetric Standard Model (MSSM) can go around both
problems.  The first order phase transition becomes a possibility
again if one of the scalar top quark is very light, $m_{\tilde{t}_R}
\lesssim 160$~GeV, despite the LEP bound.  There is also a new CP
violating phase $\Im (\mu^* M_2)$ in the chargino sector which could in
principle be order one.  See Fig.~\ref{fig:bubbles} and the caption
for the mechanism of the baryogenesis.

However, the model is getting cornered; the available
parameter space is becoming increasingly limited due to the LEP
constraints on chargino, scalar top quark and Higgs boson
\cite{Carena,Cline}.  This is because the LEP bound on the lightest
Higgs boson requires a large radiative correction, and hence a large
scalar top quark mass.  Because we need a light right-handed scalar
top quark to achieve the first order phase transition, only the
left-handed scalar top quark can be raised above TeV for this
purpose.  $\tan \beta$ also needs to be large to raise the Higgs
mass.  However, the CP violation is a relative phase between $\mu$ and
$M_2$ in the chargino mass matrix
\begin{equation}
  \label{eq:3}
  \left( \begin{array}{cc} M_2 & \sqrt{2} m_W \cos \beta\\
      \sqrt{2} m_W \sin \beta & \mu 
    \end{array} \right),
\end{equation}
and the phase becomes unphysical as $\cos \beta \rightarrow 0$ ($\tan
\beta \rightarrow \infty$).  Therefore, $\tan \beta$ cannot be large
to retain enough CP violation, causing a tension with the requirement
of heavy enough Higgs mass.  Moreover, the constraints from electric
dipole moments are quite severe if the relative phase between $\mu$
and $M_2$ is order unity.\footnote{After my talk, the constraints from
  electric dipole moments had been shown to be even more severe
  \cite{EDM}.}  What it means is that we are supposed to find a
right-handed scalar top quark, charginos ``soon'' with a large CP
violation in the mass matrix.  

\begin{figure}[t]
  \begin{center}
    \includegraphics[width=0.6\columnwidth]{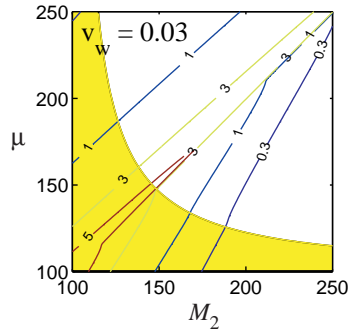}
    \caption[EWbaryogenesis]{Constraint on the MSSM chargino parameter
      space in electroweak baryogenesis \cite{Cline}.  To generate
      $\eta = 5 \times 10^{-10}$, the parameters must lie inside the
      contour labeled ``5.''  It implies light charginos.  Shaded
      region is excluded by LEP.}
    \label{fig:EWbaryogenesis}
  \end{center}
\end{figure}

It is important to ask if there is any interesting consequence of this
scenario in CP violation on B systems.  It turns out, however, that
there is no new CP violation in B systems from the relevant phase for
the electroweak baryogenesis in the MSSM
\cite{EWbaryogenesis2}.\footnote{It is quite possible that other
  models of electroweak baryogenesis lead to new CP violation in
  $B$-physics.}  The most important effect is the contribution to
$B_{d,s}$--$\overline{B}_{d,s}$ mixing.  Surprisingly, the new CP
violating phase does not appear in the diagram, and hence the mixing
amplitude has the same phase as in the Standard Model.  There is,
however 20--30\% enhancement in the mixing amplitude.  Such an
enhancement cannot be established now because of the theoretical
uncertainty in the $B$-parameter.  However lattice calculations are
expected to reduce the uncertainty down to 5\% level in the near
future, and this enhancement may be seen experimentally.  It would
require a complicated analysis.  In the case of $B_s$ mixing, the
Standard Model prediction must be improved with a better determination
of $V_{cb}$ (and hence $V_{ts}$ through unitarity) and a better
calculation of the $B$-parameter.  In the case of $B_d$ mixing, we
used to determine $V_{td}$ from the mixing.  For the purpose of
extracting a new contribution in the mixing, we have to determine
$V_{td}$ by alternative method.  By an improved determination of
$V_{ub}$ combined with measurement of angles from tree-level
processes, $V_{td}$ can in principle be determined without relying on
the $B_d$ mixing, and an enhancement in the mixing may be established.
In order to truly establish the model, we would like to see the new CP
violating phase in the chargino mass matrix.  An electron-positron
linear collider would be the best approach.

\begin{figure}[t]
  \begin{center}
    \includegraphics[width=0.6\columnwidth]{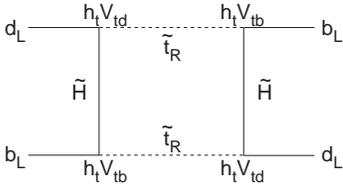}
    \caption[higgsinobox]{The most important contribution to $B$-physics
      from particles relevant for the electroweak baryogenesis in the
      MSSM.  The phase of the amplitude is exactly the same as that of
      the Standard Model diagram.  The same argument applies to $B_s$
      mixing. }
    \label{fig:higgsinobox}
  \end{center}
\end{figure}

\subsection{Leptogenesis}

\begin{figure}[t]
  \begin{center}
    \includegraphics[width=\columnwidth]{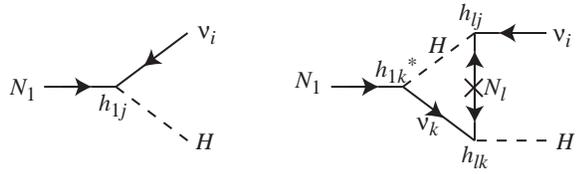}
    \caption[Ndecay]{The tree-level and one-loop diagrams of
      right-handed neutrino decay into leptons and Higgs.  The
      absorptive part in the one-loop diagram together with
      CP-violating phases in the Yukawa couplings leads to the direct
      CP violation $\Gamma (N_1 \rightarrow l H) \neq \Gamma (N_1
      \rightarrow \bar{l} H)$.}
    \label{fig:Ndecay}
  \end{center}
\end{figure}

In leptogenesis, you generate $L \neq 0$ first.  Then $L$ gets
partially converted to $B$ by the electroweak anomaly.  The question
then is how you generate $L \neq 0$.  In the original
proposal \cite{FY}, it was done by the decay of a right-handed neutrino
(say $N_1$), present in the seesaw mechanism, with a direct CP
violation.  At the tree-level, a right-handed neutrino decays equally
into $l+H$ and $\bar{l}+H^*$.  At the one-loop level, however, the
interference between diagrams shown in Fig.~\ref{fig:Ndecay} cause a
difference in the decay rates of a right-handed neutrino into leptons
and anti-leptons as
\begin{eqnarray}
  \frac
  {\Gamma(N_1 \rightarrow l_i H) - \Gamma(N_1 \rightarrow \bar{l}_i H)}
  {\Gamma(N_1 \rightarrow l_i H) + \Gamma(N_1 \rightarrow \bar{l}_i H)}
  \nonumber \\
  \simeq \frac{1}{8\pi} 
  \frac{\Im (h_{1j} h_{1k} h_{lk}^* h_{lj}^*)}{|h_{1i}|^2}
  \frac{M_1}{M_3}.
\end{eqnarray}
Once the right-handed neutrinos are produced in early Universe, their
long lifetime would allow them to decay out of equilibrium, thereby
generating an asymmetry between leptons and anti-leptons.  The lepton
asymmetry then is partially converted to baryon asymmetry thanks to
electroweak anomaly.  Much more details had been worked out in the
light of recent neutrino oscillation data and it had been shown that a
right-handed neutrino of about $10^{10}$~GeV can well account for the
cosmic baryon asymmetry from its out-of-equilibrium decay
\cite{Buchmuller}. 

There is some tension in the supersymmetric version because of
cosmological problems caused by the gravitino, which prefers a low
reheating temperature after inflation.  But a supersymmetric problem
has a supersymmetric solution.  It can be circumvented using the
superpartner of right-handed neutrino that can have a coherent
oscillation after the inflation \cite{Hamachan}.  Leptogenesis can
work.

Can we prove leptogenesis experimentally?  Lay Nam Chang, John Ellis,
Belen Gavela, Boris Kayser, and myself got together at Snowmass 2001
and discussed this question.  The short answer is unfortunately no.
There are additional CP violating phases in the heavy right-handed
neutrino sector that cannot be seen by studying the light left-handed
neutrinos.\footnote{If you believe in a certain scenario of
  supersymmetry breaking, low-energy lepton-flavor violation can carry
  information about CP violation in the right-handed neutrino sector
  \cite{Hisano}. However, such connection depends on the assumptions
  in the origin of supersymmetry breaking.}  For example, even
two-generation seesaw mechanism is enough to have CP violation that
can potentially produce lepton asymmetry, unlike the minimum of
three-generations for CP violation in neutrino oscillation.  However,
we decided that if we will see (1) electroweak baryogenesis ruled out,
(2) lepton-number violation {\it e.g.}\/ in neutrinoless double beta
decay,\footnote{There is, however, a realistic model of leptogenesis
  without lepton-number violation \cite{Dirac}.} and (3) CP violation
in the neutrino sector {\it e.g.}\/, in very long-baseline neutrino
oscillation experiment, we will probably believe it based on these
``archaeological'' evidences.

\section{Grand Unified Theories}

As we have discussed, baryogenesis may not be a good motivation for
GUTs any more.  However, there are still many reasons to consider GUTs
seriously to answer some of the big questions in the Standard Model.
For example, electric charges are quantized in the unit of $e/3$ among
quarks and leptons, while the electromagnetic $U(1)$ gauge invariance
does not require such quantization.  Similarly, the non-trivial
anomaly cancellation and seemingly random hypercharge assignments hint
at deeper organizing principles behind the quantum numbers of quarks
and leptons.  Three forces are seemingly unrelated, but they are all
based on the gauge principle.  Also philosophically, unified
description of all forces has been a dream since Einstein.  GUTs
address these questions beautifully.

The quarks and leptons are unified in GUT multiplets.  In the case of
$SU(5)$ group, {\bf 5}$^*$ multiplet includes a lepton doublet and a
right-handed down quark of three colors, and {\bf 10} multiplet
includes a quark doublet and a right-handed up quark of three colors,
and right-handed charged lepton.

\begin{figure}[t]
  \begin{center}
    \includegraphics[width=\columnwidth]{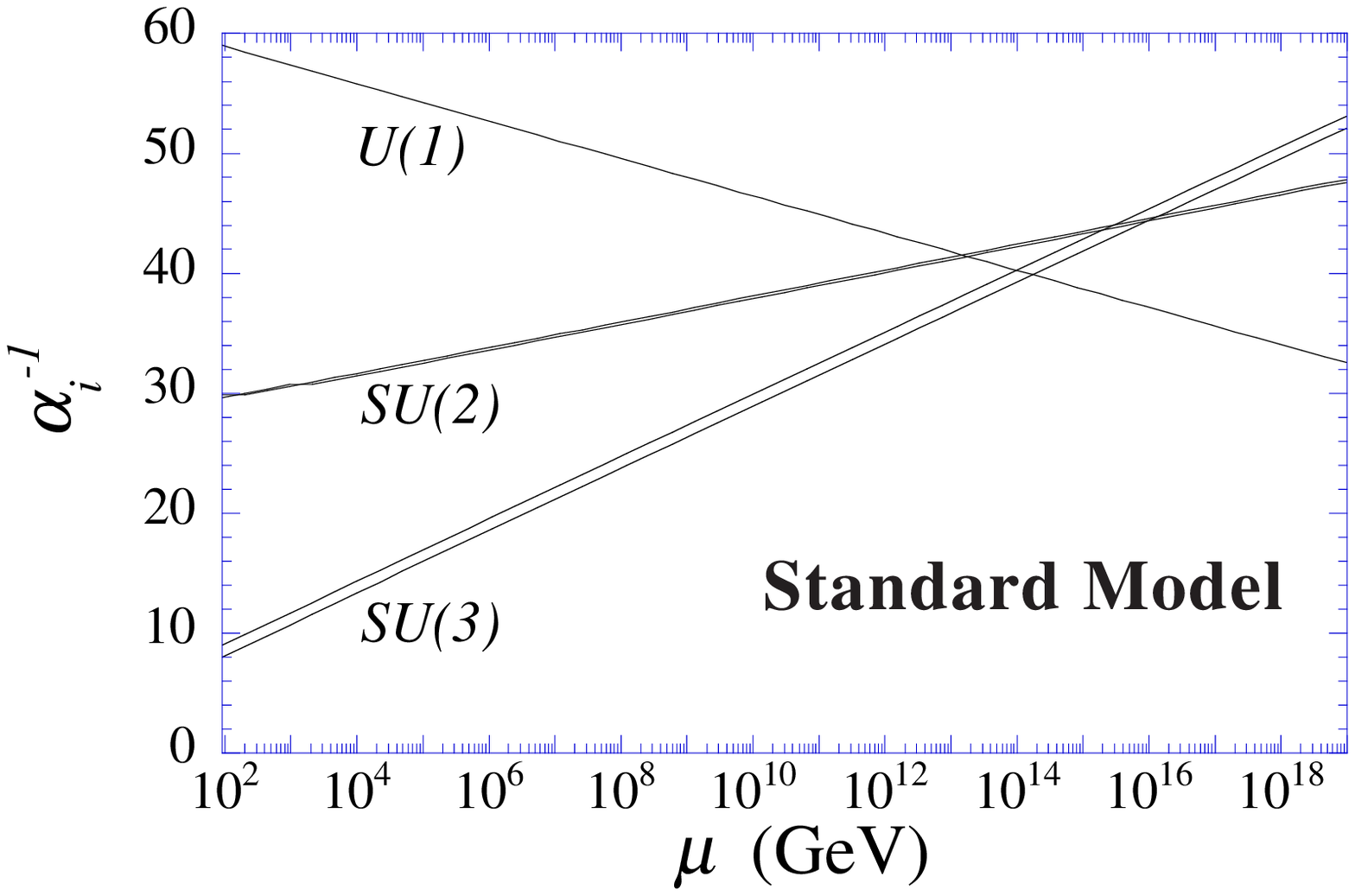}
    \includegraphics[width=\columnwidth]{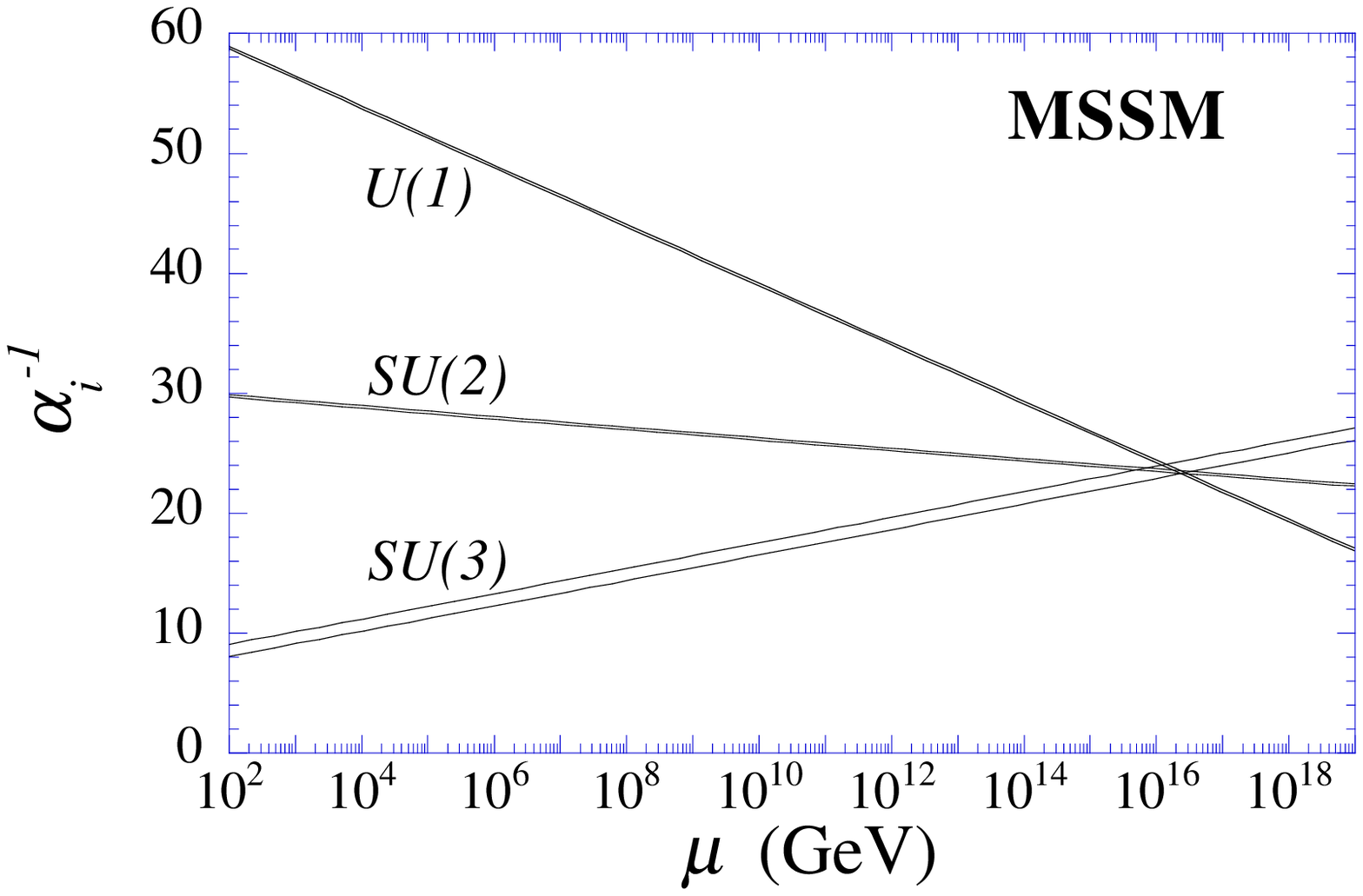}
    \caption[GUT]{The gauge coupling constants extrapolated to higher
    energies using the particle contents of the Minimal Standard Model
    and the Minimal Supersymmetric Standard Model.}
    \label{fig:GUT}
  \end{center}
\end{figure}

Phenomenologically, the original non-supersymmetric $SU(5)$ GUT no
longer works because the gauge coupling constants measured precisely
at LEP and SLC extrapolated to higher energies do not meet at a point.
However with the MSSM particle content, running of the gauge couplings
changes and they do meet within a percent level accuracy at $2 \times
10^{16}$~GeV.  Therefore, the Minimal supersymmetric $SU(5)$ GUT
appears a success phenomenologically.

\begin{figure}[t]
  \begin{center}
    \includegraphics[width=0.8\columnwidth]{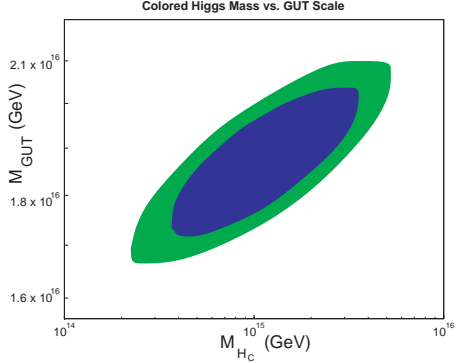}
    \caption[corr]{The GUT-scale mass parameters $M_{H_C}$ (mass of
    color-triplet Higgs that mediates the proton decay) and $M_{GUT} =
    (M_X^2 M_\Sigma)^{1/3}$ ($M_X$ is the mass of the $X$ boson, and
    $M_\Sigma$ that of the adjoint Higgs) from the requirement that
    coupling constants unify.}
    \label{fig:corr}
  \end{center}
\end{figure}

On the other hand, another prediction of GUTs than the gauge coupling
unification, namely the proton decay, has not been observed.  In fact,
the limit from SuperKamiokande $\tau (p \rightarrow \bar{\nu} K^+) >
6.7 \times 10^{32}$~yr implies the mass of the color-triplet
$SU(5)$-partner of the Higgs boson to be heavier than $7.6 \times
10^{16}$~GeV, while the coupling unification requires it to be well
below $10^{16}$~GeV (Fig.~\ref{fig:corr}).  Even with the so-called
``decoupling'' limit where first- and second-generation squarks are
assumed to be heavy, the lower bound comes down only to $5.7 \times
10^{16}$~GeV.  Therefore, the Minimal supersymmetric $SU(5)$ GUT is
now excluded \cite{proton}.

Unfortunately, the prediction of the proton lifetime is sensitive to
the ugly part of GUT model building, namely how to accommodate
observed quark and lepton masses, which the minimal model gets wrong
by a factor of three, and how to keep only the doublet Higgs light
while making color-triplet partner heavy (``triplet-doublet splitting
problem'').  Expanded particle content at the GUT scale also makes the
prediction more uncertain \cite{proton}.  Some fine-tuning can
suppress the decay rate \cite{BabuBarr}.  A recent proposal of
breaking $SU(5)$ using extra dimension on a one-dimensional orbifold
$S^1/Z_2$ eliminates this mode of proton decay \cite{orbifold}.
Therefore, one cannot say that supersymmetric GUTs are excluded.  In
other words, proton decay may be just around the corner.  It should be
remembered that proton decay is a truly unique window to physics at $>
10^{15}$~GeV scale, and it is worth pursuing anyway.  Some other
supersymmetric GUT models predict $p \rightarrow e^+ \pi^0$ mode close
to the experimental limit, including flipped $SU(5)$ model
\cite{proton} and orbifold GUT \cite{orbifold}.  

In the absence of direct signal (proton decay), it is natural to look
for other indirect effects of GUT.  It has been well studied that
quark-lepton unification causes flavor-changing effects among leptons
through top quark Yukawa coupling, giving rare phenomena such as $\mu
\rightarrow e\gamma$, $\mu \rightarrow e$ conversion at experimentally
accessible rates \cite{BH}.  Recent discovery of $\nu_\mu \rightarrow
\nu_\tau$ atmospheric neutrino oscillation with large mixing angle may
also give observable $\tau \rightarrow \mu \gamma$ etc at near future
experiments \cite{tau}.

\begin{figure}[t]
  \begin{center}
    \includegraphics[width=0.6\columnwidth]{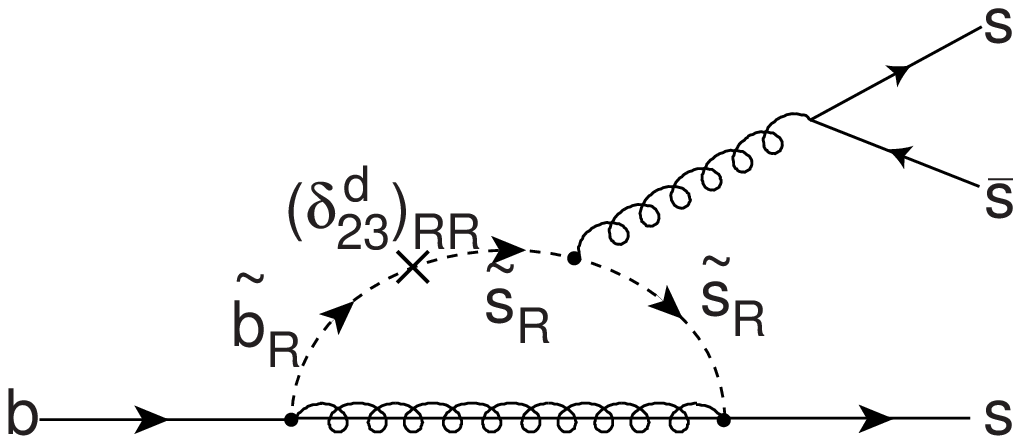}
    \includegraphics[width=0.6\columnwidth]{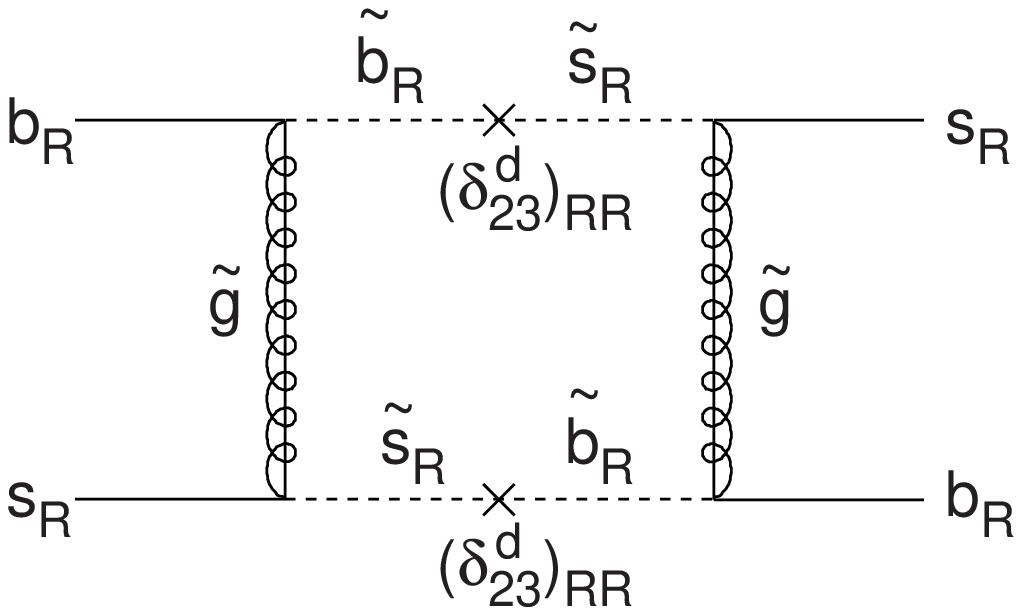}
    \caption[Darwin]{The diagrams that can contribute to direct CP
    violation in $B_d \rightarrow \phi K_s$ and $B_s$ mixing.}
    \label{fig:Darwin}
  \end{center}
\end{figure}

I would like to advertise yet another interesting signal of GUT in
$B$-physics \cite{Darwin}.  Take the large mixing between $\nu_\mu$
and $\nu_\tau$.  In $SU(5)$ GUT, they are in the same multiplets as
$s_R$ and $b_R$, respectively.  Therefore, it is natural to expect a
large mixing between $s_R$ and $b_R$.  Such mixing drops out
completely from CKM matrix because it keeps track of mixing only among
left-handed quarks that participate in the charged-current weak
interaction.  However, in presence of supersymmetry, a large mixing
between $\tilde{s}_R$ and $\tilde{b}_R$ would lead to observable
effects.  For example, supersymmetric contribution to $B_s$ mixing may
well be comparable to the standard model contribution even with
squarks at 1~TeV.  Supersymmetric penguin contribution to $b
\rightarrow s \bar{s} s$ allows direct CP violation and may give
different ``$\sin 2\beta$'' in $B_d \rightarrow \phi K_s$ from that in
$B_d \rightarrow J/\psi K_s$.\footnote{Recently both BABAR and BELLE
  collaborations announced discrepancy between two measurements of
  ``$\sin 2\beta$'' at two standard deviation each \cite{phiKs}.}

\section{Models of Flavor}

The most pressing question on the origin of flavor is what
distinguishes flavor.  Three generations of quarks and leptons have
exactly the same quantum numbers.  But then, how come that they have
so different masses and small mixings?  When we learn quantum
mechanics, we see that states with similar quantum numbers have
similar energies, and they mix greatly.  Due to some reason, three
generations do not follow this common wisdom.  Why?

One way to answer this question, developed very actively in the past
few years, is the concept of explicitly broken flavor symmetry.  The
idea is simple: there must a new (but hidden) quantum number that
distinguishes different flavors.  Because of the difference in the
flavor quantum number, particles in different generations have very
different masses and mix little.  According to N\"other's theorem, a
quantum number means a symmetry, and hence flavor symmetry.  The
flavor symmetry must allow top quark Yukawa coupling because it is a
coupling of order unity which we consider natural.  Other Yukawa
couplings are very small.  They are forbidden by the flavor symmetry,
and are generated only when the symmetry breaks.  The smallness of the
Yukawa couplings are controlled by the smallness of symmetry
breaking.  The mixings are also forbidden by the flavor symmetry, and
are induced by the symmetry breaking.

To make the discussion more concrete, let us study the following
$SU(5)$-like flavor quantum number assignment.  I choose a simple
$U(1)$ charge, and I assign \cite{Haba}
\begin{eqnarray}
  \label{eq:4}
  \begin{array}{ccc}
    {\bf 10}_1 (+2), & {\bf 10}_2 (+1) & {\bf 10}_3 (0) \\
    {\bf 5}_1 (+1), & {\bf 5}_2 (+1), & {\bf 5}_3 (+1).
  \end{array}
  \label{eq:charges}
\end{eqnarray}
The subscripts refer to the generations, and the charges shown in
parentheses.  With this assignment, the only allowed Yukawa coupling
is the one for the top quark.  Other Yukawa couplings are forbidden.
Next I assume that the $U(1)$ symmetry is broken by a small parameter
$\epsilon \sim 0.05$ that carries the charge $-1$.  Then it becomes
possible to fill in all elements of the mass matrices,
\begin{eqnarray}
  M_u &\sim& \left( 
    \begin{array}{ccc}
      \epsilon^4 & \epsilon^3 & \epsilon^2 \\
      \epsilon^3 & \epsilon^2 & \epsilon \\
      \epsilon^2 & \epsilon & 1
    \end{array} \right), \nonumber \\
  M_d &\sim& \left( 
    \begin{array}{ccc}
      \epsilon^3 & \epsilon^3 & \epsilon^3 \\
      \epsilon^2 & \epsilon^2 & \epsilon^2 \\
      \epsilon & \epsilon & \epsilon
    \end{array} \right),
  \nonumber \\
  M_l &\sim& \left( 
    \begin{array}{ccc}
      \epsilon^3 & \epsilon^2 & \epsilon \\
      \epsilon^3 & \epsilon^2 & \epsilon \\
      \epsilon^3 & \epsilon^2 & \epsilon
    \end{array} \right).
\end{eqnarray}
The symbol $\sim$ emphasizes that we cannot predict precise numbers
based on this simple idea, but only the order of magnitudes suppressed
by powers of $\epsilon$.  A simple approximate prediction out of this
charge assignment is that
\begin{eqnarray}
  \lefteqn{
    m_u : m_c : m_t \sim m_d^2 : m_s^2 : m_b^2 
  }\nonumber \\
  & & \sim m_e^2 : m_\mu^2 : m_\tau^2
  \sim \epsilon^4 : \epsilon^2 : 1,
\end{eqnarray}
which is phenomenologically acceptable.  Note that up quarks are doubly
hierarchical compared to the down quarks and charged leptons.  Mixing
angles are also predicted, $V_{cb} \sim V_{ts} \sim \epsilon$, $V_{ub}
\sim V_{td} \sim \epsilon^2$, which work quite well, and $V_{us} \sim
V_{cd} \sim \epsilon$ is a little bit too small but not crazy.

What about neutrinos?  Indeed, recent data on neutrino oscillations
have shed considerable insight into the flavor symmetry.  The MNS
matrix suggested by atmospheric, solar (LMA solution), and reactor
data has the form
\begin{eqnarray}
  \lefteqn{
    (e \ \mu \ \tau)
    \left( 
      \begin{array}{ccc}
        {\it large} & {\it large} & {\it smallish} \\
        {\it large} & {\it large} & {\it large} \\
        {\it large} & {\it large} & {\it large}
      \end{array} \right)
    \left(
      \begin{array}{c}
        \nu_1 \\ \nu_2 \\ \nu_3
      \end{array} \right),
  } \nonumber \\
  & & \frac{\Delta m^2_{12}}{\Delta m^2_{23}} = 0.01\mbox{--}0.2.
\end{eqnarray}
The mass hierarchy is not very large, especially after taking square
roots.  All angles are large except $|U_{e3}| \lesssim 0.15$, but even
this constraint is not particularly strong compared to $U_{e2} \sim
0.4$, $U_{\mu 3} \sim 0.7$.  It has been a big surprise to all of us
that the pattern is so different from quarks and charged leptons.

I actually find the neutrino masses and mixings very natural.  In view
of the question I posed earlier, if three quantum mechanical states
share the same quantum numbers, their energies (masses) are expected
to be similar, and their mixings unsuppressed.  We have been so much
used to hierarchical masses and small mixings over many decades, but
what is surprising is not the neutrinos but other quarks and leptons
rather!  I view the observed pattern of neutrino masses and mixings a
great confirmation of our naive intuition.  

In terms of flavor quantum numbers, all we need to do then is to
assign the same quantum numbers to three generations of neutrinos.
Indeed, the charge assignment (\ref{eq:charges}) was motivated by this
requirement to obtain
\begin{equation}
  M_\nu \sim \left( 
    \begin{array}{ccc}
      1 & 1 & 1 \\
      1 & 1 & 1 \\
      1 & 1 & 1
    \end{array} \right).
\end{equation}
Here, the overall suppression of $\epsilon^2$ is dropped because the
overall mass scale of neutrino masses is probably determined by other
physics such as seesaw mechanism.  

But you may wonder if such an argument may only produce largish
mixing angles, but never dramatically maximal mixing as in atmospheric
neutrino oscillations.  It turns out that a maximal mixing is in a
sense the most natural angle.  Using a simple Monte Carlo of random
neutrino mass matrices, you can see that there is a peak in the
distribution at $\sin^2 2\theta_{23} = 1$.  Actually, you can
understand this distribution based on purely group-theoretic
consideration.  The unique invariant measure (Haar measure) of $SU(3)$
MNS matrix gives this distribution.

\begin{figure}[t]
  \begin{center}
    \includegraphics[width=0.6\columnwidth]{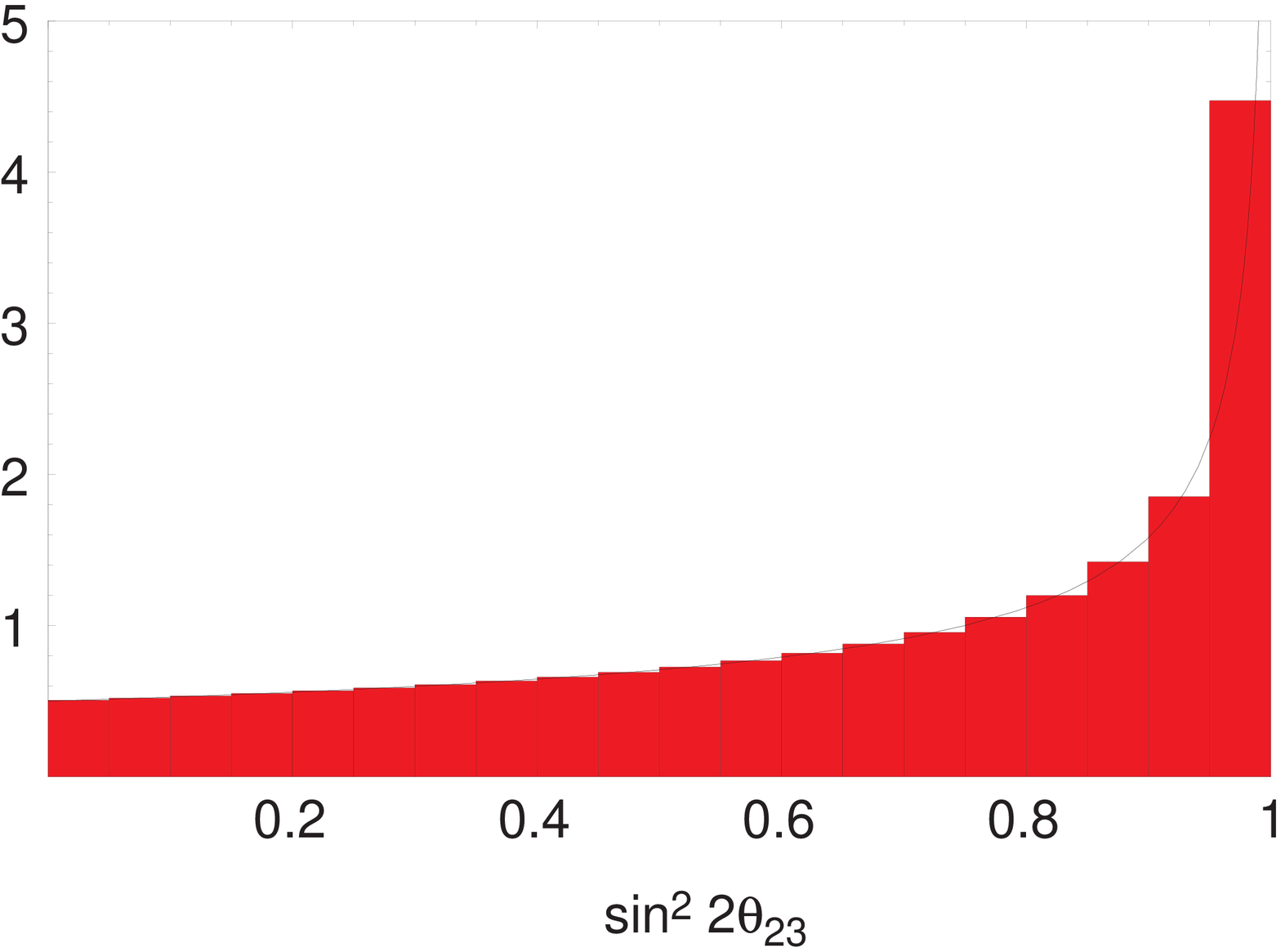}
    \includegraphics[width=0.6\columnwidth]{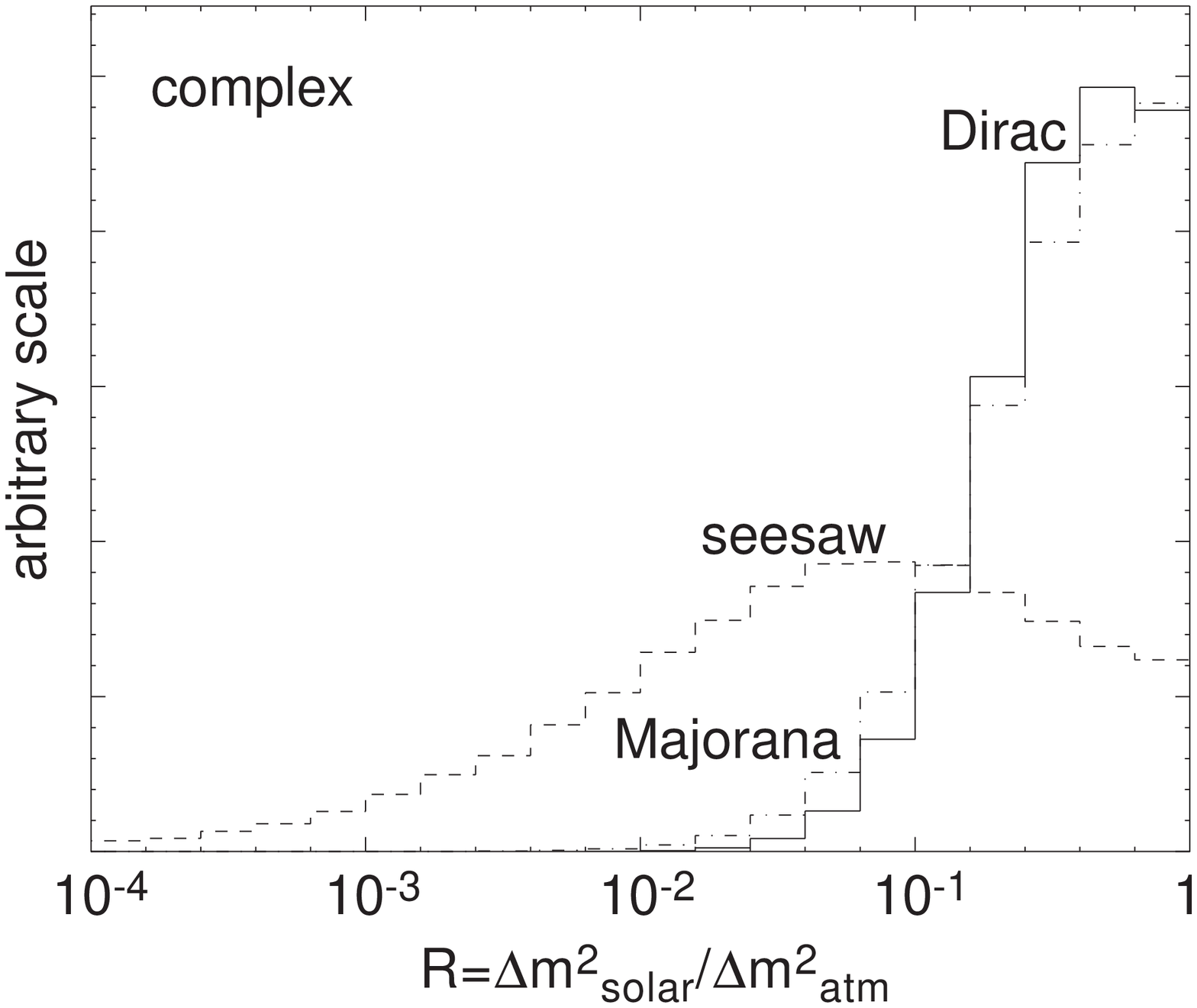}
    \caption[anarchy]{The distributions in $\sin^2 2\theta_{23}$ and
    $\Delta m^2_{12}/\Delta m^2_{23}$ from randomly generated neutrino
    mass matrices \cite{Haba}.}
    \label{fig:anarchy}
  \end{center}
\end{figure}

We called this simple fact ``anarchy'' \cite{anarchy,Haba}.  Because
there is no quantum number to distinguish three generations of
neutrinos, neutrino mass matrix lacks any particular structure.  That
alone explains large mixing angles and small hierarchy.  Even though
$\sin^2 2\theta_{13}$ appears indeed small compared to the obtained
distribution, if you get three quantities $\sin^2 2\theta_{23}$,
$\sin^2 2\theta_{12}$, and $\Delta m^2_{12}/\Delta m^2_{23}$ right
with one outlier, I find it perfectly reasonable.  The anarchy then
predicts the LMA solution, testable at KamLAND, and $\theta_{13}$ just
below the bound.  CP violation is also $O(1)$.  This is the prefect
scenario for long-baseline neutrino oscillation experiments.

\section{Conclusion}

Flavor physics is going through an amazing period.  Just to name a few
important points I covered in this talk, (1) Electroweak baryogenesis
is getting cornered, (2) Leptogenesis is gaining momentum, and (3)
Neutrinos provide new insight into the origin of flavor.  The good news
is that we will obtain more data, including rare decays and possible
deviations from the Standard Model, to hopefully pin down the flavor
symmetry and eventually its dynamical origin.

\section*{Acknowledgements}

I thank the organizers for inviting me to the exciting workshop and
also for the patience waiting for my manuscript.  This work was
supported in part by the DOE Contract DE-AC03-76SF00098 and in part by
the NSF grant PHY-0098840.

\end{document}